\documentclass{article}

\usepackage{amssymb}
\usepackage{amsmath}
\usepackage{float}
\usepackage{cite}
\usepackage{graphics}
\usepackage{epsfig}
\usepackage{subfigure}

\usepackage{graphicx}

\begin{document}
\noindent \noindent \textbf{} \textbf{\centerline{\LARGE $f(R,T)$ Gravity Model Behaving as a Dark Energy Source}}\\

\noindent \textbf{}

\noindent \textbf{}

\noindent \noindent \textbf{} \textbf{\centerline{$^1$Pheiroijam Suranjoy Singh, $^2$Kangujam Priyokumar Singh }}\\

\noindent \noindent {\centerline{  $^1$$^,$$^2$Department of Mathematical Sciences,}}

\noindent \noindent {\centerline{  Bodoland University, Kokrajhar, Assam-783370, India}}

\noindent \noindent {\centerline{  $^2$Department of Mathematics,}}

\noindent \noindent {\centerline{  Manipur University, Imphal, Manipur-795003, India}}\\

\noindent \noindent {\centerline{$^1$ E-mail : surphei@yahoo.com}}

\noindent \noindent {\centerline{ $^2$E-mail : pk\textunderscore{mathematics@yahoo.co.in}}}\\

\noindent \textbf{}

\noindent

\noindent

\begin{abstract}
Within the limits of the present cosmological observations in $f(R, T)$ gravity theory, we have analyzed a spherically symmetric space-time in 5D setting. The field equations have been carefully studied considering reasonable cosmological assumptions to obtain exact solutions. We have obtained an isotropic model universe undergoing super-exponential expansion. It is predicted that the model universe behaves like a dark energy (vacuum energy) model. In the present scenario, the model evolves with a slow and uniform change of shape. It is observed that the universe is close to or nearly flat. The model is free from initial singularity and is predicted to approach the de-Sitter phase dominated by vacuum energy or cosmological constant in the finite-time future. A comprehensive discussion on the cosmological parameters obtained in view of the recent studies is presented in detail with graphs.  
\end{abstract}


\footnote{Preprint of an article published in [New Astronomy, DOI: 10.1016/j.newast.2020.101542]
[Elsevier] [Journal URL: www.sciencedirect.com/journal/new-astronomy]
}

\section{Introduction}

The ambiguous dark energy (DE) has been regarded as one of the most tantalizing topics in cosmology since its profound discovery in 1998 \cite{1,2}. It is considered to be the reason behind the late time expanding universe at an expedited rate due to its huge negative pressure with repulsive gravitation. It is uniformly permeated throughout the space and vary slowly or almost consistent with time \cite{3,4,5,6}. Cosmologists all over the map have conducted a series of studies with the aim of hunting its origin and are still scrabbling for a perfect answer. Some worth mentioning such studies that have not escaped our notice in the recent years are briefly discussed below.

In \cite{7}, the authors assert that emergent D-instanton might indicate us a hint to the root of DE. A cosmological model associated with an antineutrino star is constructed by \cite{8} in order to search the origin of DE. In \cite{9}, the author presents an explanation for DE with pure quantum mechanical method. In \cite{10}, the investigation of the twenty years old history of DE and the current status can be seen. The authors in \cite{11} study the evolution of the DE using a non-parametric Bayesian approach in the light of the latest observation. In \cite{12}, the author claims that vacuum condensate can provide us the origin of DE. According to \cite{13}, DE  is originated from the violation of energy conservation. A unified dark fluid is obtained as a source of DE by \cite{14}. The presence of particle with imaginary energy density can lead us to the source of DE \cite{15}. The explanation of a physical mechanism as a source of DE is presented by \cite{16}. Lastly, in \cite{17}, DE evolves as a result of the condensation of fermions formed during the early evolution.    

It is an obvious fact that the universe is dominated by the cryptic DE with negative pressure and positive energy density \cite{5,18,19,20,21,22,23,24}. This qualifies DE a completely irony of nature as the dominating component is also the least explored. As against the positive energy density condition, it is fascinating to see many authors introducing the concept of possibility of negative energy density (NED) with convincig arguments in support. In \cite{25}, the authors discuss NED where models evolve with a bounce. The authors continued that there might be bounces in the future too. The discussion of negative vacuum energy density in Rainbow Gravity can be seen in \cite{26} . In \cite{27}, we can witness, under certain conditions, a repelling negative gravitational pressure with NED. Further, we can find a repelling negative phantom energy with NED. In \cite{28}, the author claims that the universe evolves by inflation when the coupled fluid has NED in the initial epoch. An accelerating universe with NED is studied by \cite{29}. In \cite{30}, we can find an explaination of energy density with negative value with equation of state parameter (EoS) $\omega<-1$. The author in \cite{5} predicts that NED is possible only if the DE is in the form of vacuum energy. In \cite{31}, the investigation of models which evolved with NED in the infinite past can be found. According to \cite{32}, the introduction of quantized matter field with NED to energy momentum tensor might by pass cosmological singularity. Besides defying the energy conditions of GR, NED also disobeys the second law of thermodynamics \cite{33}. However, the condition should be solely obeyed on a large scale or on a mean calculation, thereby neglecting the probable violation on a small scale or for a short duration, in relativity\cite{34,35,36,37,38,39,40,41,42}. Hence, in the initial epoch, if there were circumstance of defiance for a short duration measured against the present age of $13.830\pm0.037$ Gyr estimated by the latest Planck 2018 result \cite{43}, it will remain as an important part in the course of evolution.     

In the present cosmology, authors prefer to opt alternate or modified theories of gravitation in order to precisely understand the underlying mechanism of the late time expedited expansion of the universe. One such well appreciated modified theory is the $f(R,T)$ gravity introduced by \cite{44} in which the gravitational Lagrangian is represented by an arbitrary function of the Ricci scalar $R$ and the trace $T$ of the energy-momentum tensor. In the past few years, this theory has captivated many cosmologists and theoretical physicists as it presents natural gravitational substitutes to DE \cite{45}. Recently, \cite{46} studies the theory and predicts the conditions to obtain expanding universe in the absence of any dark component. In \cite{47}, the authors investigate a mixture of barotropic fluid and DE in $f(R,T)$ gravity where the model evolves from the Einstein static era and approaches $\Lambda$CDM. In \cite{48}, the study of cosmological dynamics of DE within the theory can be seen. \cite{49} study a modified holographic Ricci DE model in the theory obtaining a singularity free model. The authors in \cite{50} investigate $f(R,T)$ gravity discussing future singularities in DE dominated universe. In \cite{51}, we can find a discussion of new holographic DE model in $f(R,T)$ gravity thereby obtaining $\Lambda$CDM in the late times. In \cite{52}, the examination of ghost DE model within the theory can be seen, predicting model behaving as phantom or quintessence like nature. The investigation of cosmological models within the theory without DE is observed in \cite{53} . The authors in \cite{54} and \cite{55} study the relation of the theory with DE. Houndjo and Piattella \cite{56} present a reconstruction of  the theory from holographic DE. The study cosmological model in $f(R,T)$ gravity obtaining DE induced cosmic acceleration can be seen in \cite{57}. Zubair et al. \cite{58} discuss Bianchi space-time within the theory with time-dependent deceleration parameter. Ahmed et al. \cite{59} investigate model in which the cosmological constant is considered as a function of $T$. The authors in \cite{60} discuss a higher dimensional anisotropic DE model within the theory obtaining the EoS parameter $\omega=-1$. Jamil et al. \cite{61} construct models within the theory asserting that dust fluid leads to $\Lambda$CDM. Houndjo \cite{62} predicts a model in $f(R,T)$ gravity that transit from matter dominated to accelerating phase. From these worth appreciating studies, it won't be a wrong guess to sum up that there must be some sort of hidden correspondence between the pair of DE and $f(R,T)$ gravity. Consequently, in this work, we will try to find out if $f(R,T)$ itself behaves as a DE source.

The possibility of space-time possessing with more than 4D has fascinated many authors. In the recent years, there has been a trend of preferring higher dimensional space-time to study cosmology. Higher dimensional model was introduced by \cite{63} and \cite{64} in an effort to unify gravity with electromagnetism. Higher dimensional model can be regarded as a tool to illustrate the late time expedited expanding paradigm \cite{65}. Investigation of higher dimensional space-time can be regarded as a task of paramount importance as the universe might have come across a higher dimensional era during the initial epoch \cite{66}.  Marciano \cite{67} asserts that the detection of a time varying fundamental constants can possibly show us the proof for extra dimensions. According to \cite{68} and \cite{69}, extra dimensions generate huge amount of entropy which gives possible solution to flatness and horizon problem. Since we are living in a 4D space-time, the hidden extra dimension in 5D is highly likely to be associated with the invisible DM and DE \cite{70}. 

Keeping in mind the above notable works by different authors, we have analysed a spherically symmetric metric in 5D setting within the framework of $f(R,T)$ gravity with focus to predict a possible source of DE. Here, we observe the field equations with due consideration of reasonable cosmological assumptions within the limit of the present cosmological scenario. The paper has been structured into sections. In Sect. 2, in addition to obtaining the solutions of the field equations, the cosmological parameters are also solved. In Sect. 3, the physical and kinematical aspects of our model are discussed with graphs. Considering everything, a closing remark is presented in Sect. 4.

\section{The field equations of $f(R, T)$ gravity theory}

The action of $f(R,T)$ gravity theory is given by 

\begin{equation}
S=\int  \left(\frac{1}{16\pi } f(R,T)+\mathcal{L}_{m} \right)\sqrt{-g} d^{4} x 
\end{equation}

\noindent where $g\equiv det(g_{ij})$, $f$ is an arbitrary function of the Ricci scalar $R=R(g)$ and the trace  $T=g^{ij} T_{ij}$ of the energy-momentum tensor of matter $T_{ij} $ defined by \cite{71} as
\begin{equation}
T_{ij} =-\frac{2}{\sqrt{-g} } \frac{\delta \left(\sqrt{-g} \mathcal{L}_{m} \right)}{\delta \, g^{ij} }           
\end{equation}

Here, the matter Lagrangian density $\mathcal{L}_{m} $ is assumed to rely solely on $g_{ij} $ so that we obtain

\begin{equation}
T_{ij} =g_{ij} \mathcal{L}_{m} -2\frac{\partial \mathcal{L}_{m} }{\partial g^{ij} }      
\end{equation}

The action $S$ is varied w.r.t. the metric tensor $g^{ij} $ and hence, the field equations of $f(R,T)$ gravity is given by
\begin{equation}
f_{R}\left(R,T\right)R_{ij}-\frac{1}{2}f\left(R,T\right)g_{ij}+\left(g_{ij} \square-\nabla_{i} \nabla_{j}\right)f_{R}\left(R,T\right) = 8 \pi T_{ij} -f_{T}\left(R,T\right)T_{ij}-f_{T}\left(R,T\right)\theta _{ij}  
\end{equation}
\noindent where   
\begin{equation}
\theta _{ij} =-2 \,T_{ij} +g_{ij} \mathcal{L}_{m} -2g^{lk} \frac{\partial ^{2} \mathcal{L}_{m} }{\partial g^{ij} \partial g^{lk} } \end{equation}  

Here, the subscripts appearing in $f$ represent the partial derivative w.r.t. $R$ or $T$ and $ \square\equiv \nabla^{i}\nabla_{i}, \nabla_{i}$ being the covariant derivative.\\

With $\rho $ and $p$ respectively representing the energy density and pressure such that the five velocity $u^{i} $ satisfies $u^{i} u_{i} =1$ and $u^{i} \nabla _{j} u_{i} =0$, we opt to use the perfect fluid energy-momentum tensor of the form 
\begin{equation}
T_{ij} =\left(p+\rho \right)u_{i} u_{j} -pg_{ij}          
\end{equation}  

We assume that $\mathcal{L}_{m}=-p$ so that equation (5) is reduced to 
\begin{equation}
\theta _{ij} =-2T_{ij} -pg_{ij}            
\end{equation}  
In general, the field equations of $f(R,T)$ gravity also rely on the physical aspect of the matter field and consequently, there exists three classes of field equations as follows
\begin{equation}
f(R,T)=\left\{
\begin{array}{ll}
      R+2f(T) \\
      f_{1} (R)+f_{2} (T) \\
      f_{1} (R)+f_{2} (R)f_{3} (T) \\
\end{array} 
\right.       
\end{equation} 
Our study will be dealing with the class $f(R,T)=R+2f(T)$, where $f(T)$ represents an arbitrary function so that the field equations of the modified theory is be reduced to

\begin{equation}
R_{ij} -\frac{1}{2} Rg_{ij} =8\pi \, T_{ij} +2f^{\prime} (T)\, T_{ij} +\left\{\, 2p\, f^{\prime} (T)+f(T)\, \right\}\, g_{ij\, }        
\end{equation}

\noindent where the prime indicates differentiation w.r.t. $T$ and we assume that $f(T)=\lambda T,$ where $\lambda $ is an arbitrary constant.

\section{Formulation of the problem and solutions}

The five-dimensional spherically symmetric metric is given by
\begin{equation}
ds^{2}=dt^{2}-e^{\mu}\left(dr^{2}+r^{2}d\theta^{2}+r^{2}\sin^{2}\theta d\phi^{2}\right)-e^{\delta}d\upsilon^{2}                 
\end{equation}

\noindent where $\mu=\mu(t)$ and $\delta=\delta(t) $ are cosmic scale factors.\\

Now, using co-moving co-ordinates, the surviving field equations are obtained as follows

\begin{equation} 
-\frac{3}{4} \left(\dot{\mu }^{2} +\dot{\mu }\dot{\delta }\right)=\left(8\pi +3\lambda \right)\rho -2p\lambda                    
\end{equation}
\begin{equation}
\ddot{\mu }+\frac{3}{4} \dot{\mu }^{2} +\frac{\ddot{\delta }}{2} +\frac{\dot{\delta }^{2} }{4} +\frac{\dot{\mu }\dot{\delta }}{2} =\left(8\pi +4\lambda \right)p-\lambda \rho          
\end{equation}
\begin{equation}
\frac{3}{2} \left(\ddot{\mu }+\dot{\mu }^{2} \right)=\left(8\pi +4\lambda \right)p-\lambda \rho           
\end{equation}

\noindent where an overhead dot indicates differentiation w.r.t. $t$. \\

From Eqs. (12) and (13), the expressions for the cosmic scale factors are obtained as
\begin{equation}
\mu =a-3\log \left(2\left(k-3t\right)\right)         
\end{equation}
\begin{equation}
\delta =b-3\log \left(2\left(k-3t\right)\right)       
\end{equation}

\noindent where $a, b, k$ are arbitrary constants.\\

Now, the expressions for spatial volume $v$, scalar expansion $\theta$, Hubble's parameter $H$, deceleration parameter $q$, shear scalar $\sigma^{2}$ and anisotropic parameter $A_h$ are obtained as follows.\\
\begin{equation}
v=e^{\frac{3a+b}{2} } \left(2\left(k-3t\right)\right)^{-6}      
\end{equation} 
\begin{equation}
\theta =18\left(k-3t\right)^{-1}
\end{equation}
\begin{equation}
H=\frac{9}{2} \left(k-3t\right)^{-1}
\end{equation}
\begin{equation}
q=-1.7
\end{equation}
\begin{equation}
\sigma ^{2} =\left(\frac{27-2\left(k-3t\right)}{18\left(k-3t\right)} \right)^{2}
\end{equation}
\begin{equation}
A_{h} =0
\end{equation}

From Eqs. (11) and (13), the expressions for the pressure $p$ and  the energy density $\rho$ of the model universe are respectively obtained as
\begin{equation}
p=\frac{243\lambda -324\left(8\pi +3\lambda \right)}{4\left(k-3t\right)^{2} \left(-5\lambda^{2}-32\pi^{2}-28\pi\lambda\right)}         
\end{equation}
\begin{equation}
\rho=\frac{\left(8\pi +4\lambda\right)}{\lambda}\left(\frac{243\lambda-324\left(8\pi +3\lambda\right)}{4\left(k-3t\right)^{2}\left(-5\lambda^{2}-32\pi^{2}-28\pi\lambda\right)}\right)\\-\frac{162}{\lambda\left(k-3t\right)^{2}}     
\end{equation}
The expression for the scalar curvature $R$ is obtained as  
\begin{equation}
R=\frac{513}{\left(k-3t\right)^{2}}
\end{equation}

\section{Discussions}
For convenience sake and to obtain realistic results, specific values of the arbitrary constants involved are chosen i.e., $a=b=1, k=15$ and $\lambda=-5.06911$ and $-12.5856$. The graphs of the cosmological parameters w.r.t. cosmic time $t$ are presented with the detailed discussion in view of the latest observations.
\begin{figure}[H]
\centering
\includegraphics[width=0.5\textwidth]{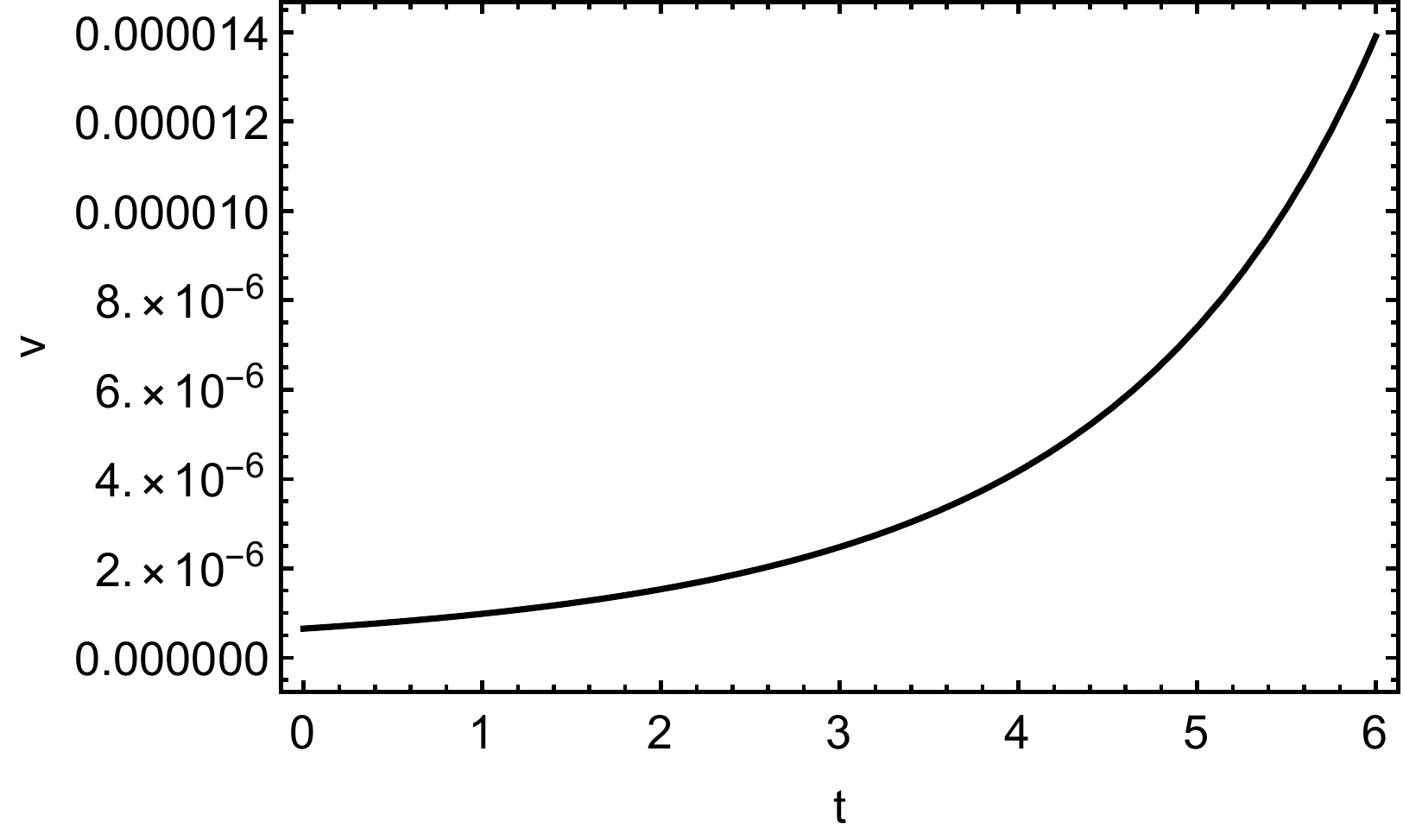}
\caption{Variation of the spatial volume $v$ with \textit{t} when $a=b=1, k=15$.}
\label{fig:2}       
\end{figure}
\begin{figure}[H]
\centering
\includegraphics[width=0.5\textwidth]{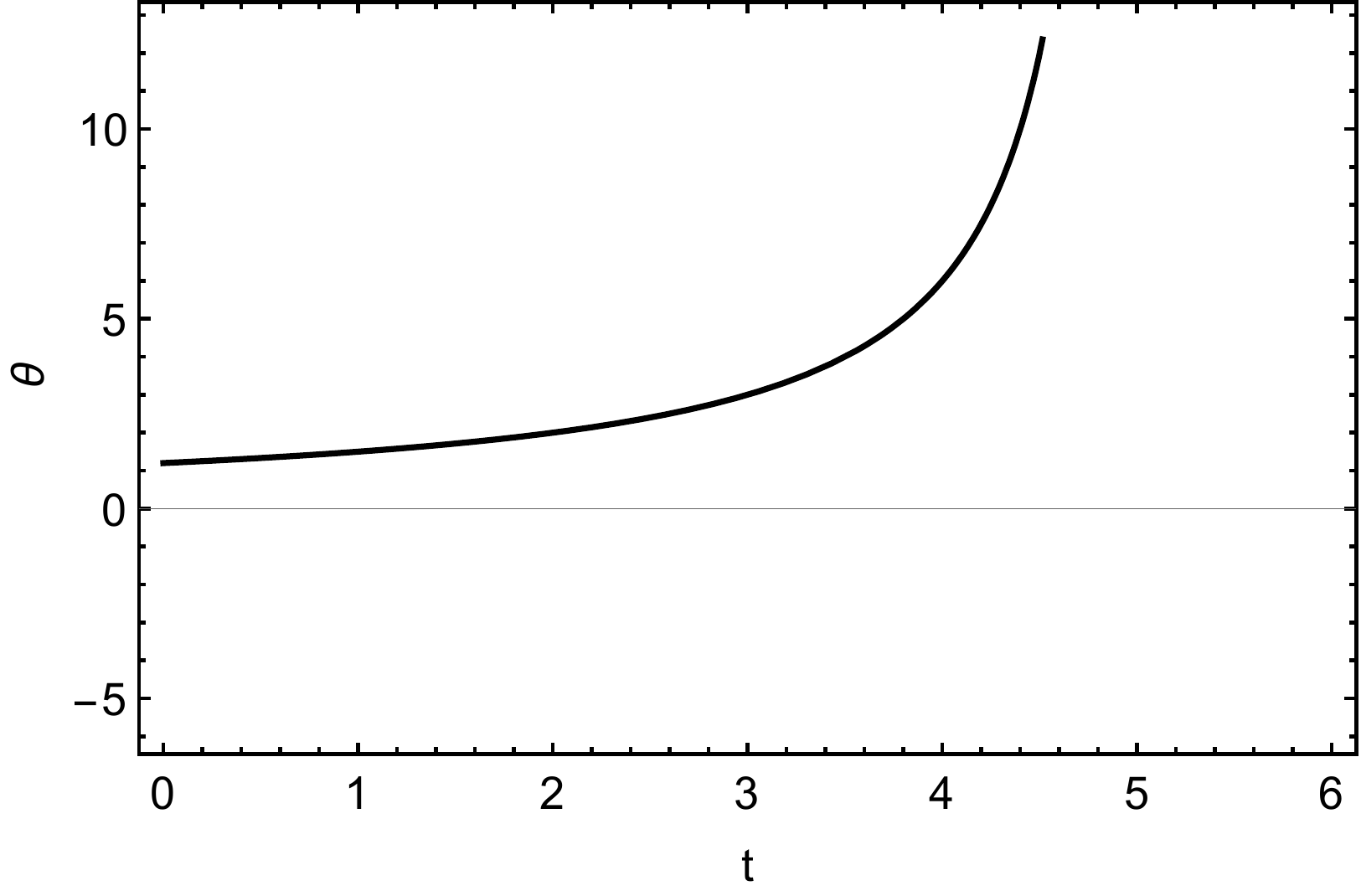}
\caption{Variation of the expansion scalar $\theta$ with \textit{t} when $a=b=1, k=15$.}
\label{fig:2}       
\end{figure}
Fig. 1 and Fig. 2 can be regarded as the perfect evidences for the present spatial expansion at an expedited rate. When $t\rightarrow0$, $v$ and other related parameters are constants ($\neq0$), implying that the model universe doesn't evolve from an initial singularity. 
\begin{figure}[H]
\centering
\subfigure[$p$ and $\rho$ with $t$ when $\lambda=-5.06911$.]{
   \includegraphics[width=0.45\textwidth]{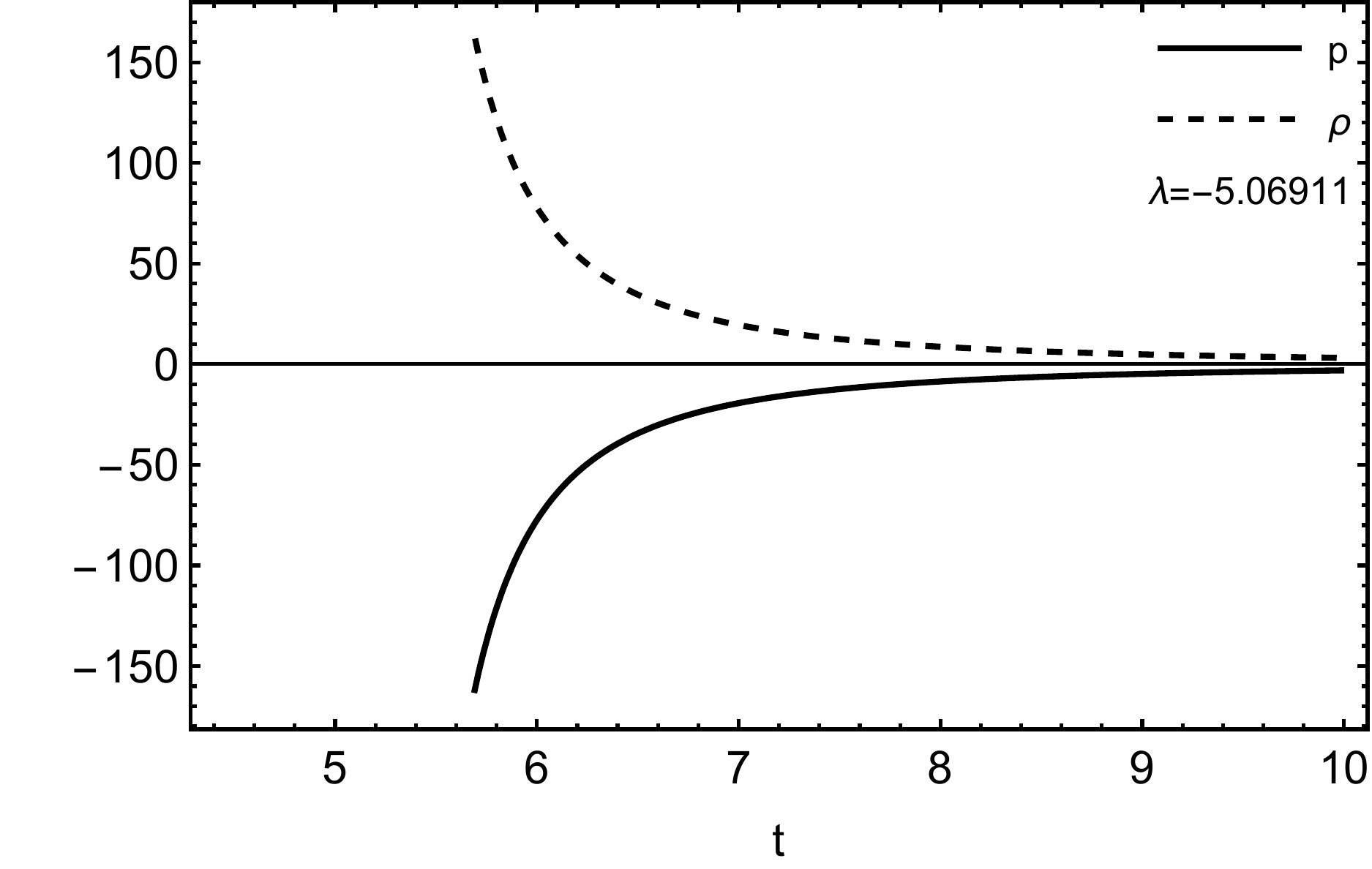}
    \label{fig:subfig1}
}
\subfigure[$p$ and $\rho$ with $t$ when $\lambda=-12.5856$.]{
   \includegraphics[width=0.45\textwidth]{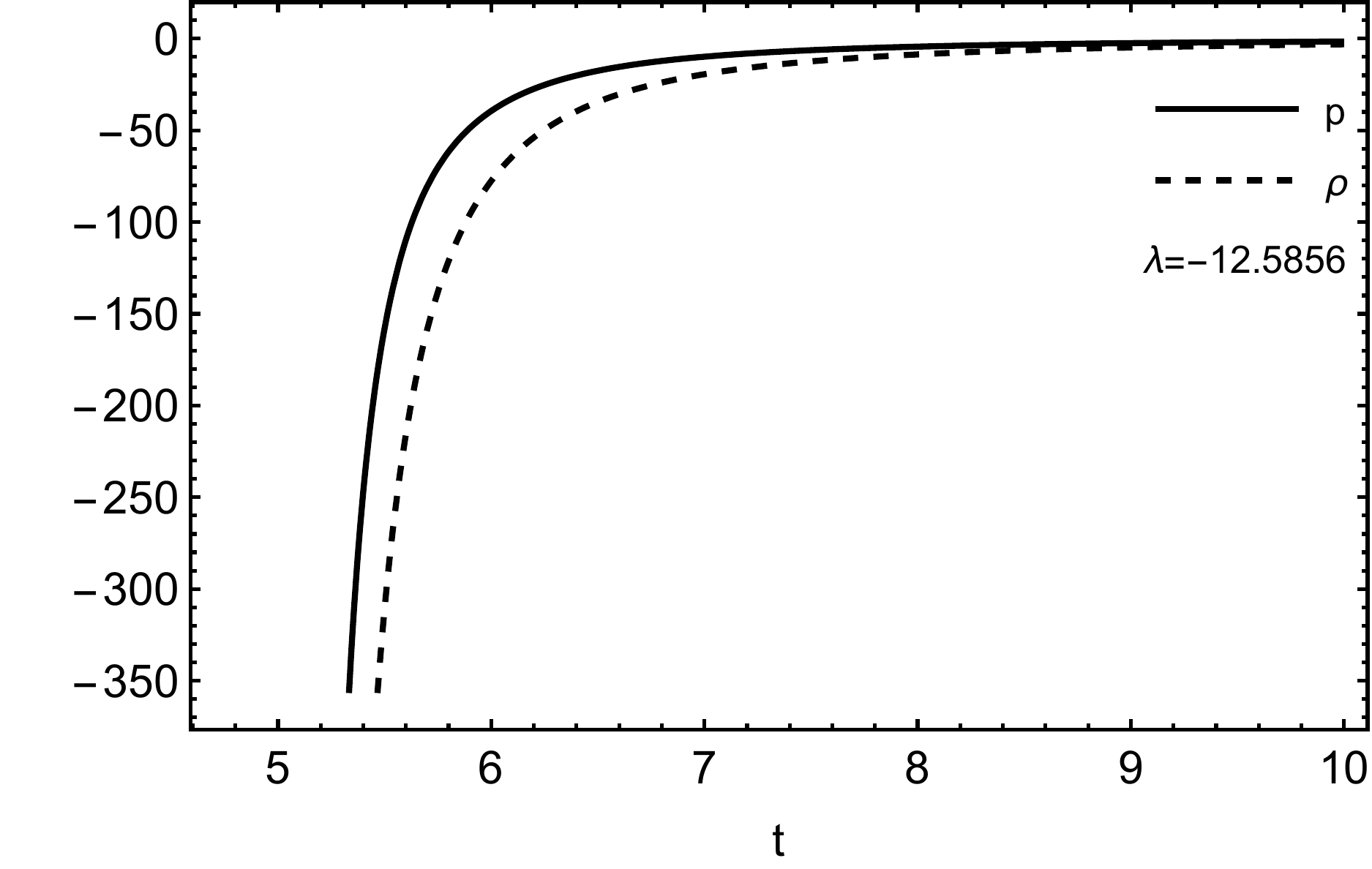}
    \label{fig:subfig2}
}
\caption[]{Variation of pressure $p$ and energy density $\rho$ when $a=b=1, k=15, \lambda=-5.06911$ and $-12.5856$.}
\label{fig:subfigureExample}
\end{figure}
\begin{figure}[H]
\centering
\includegraphics[width=0.4\textwidth]{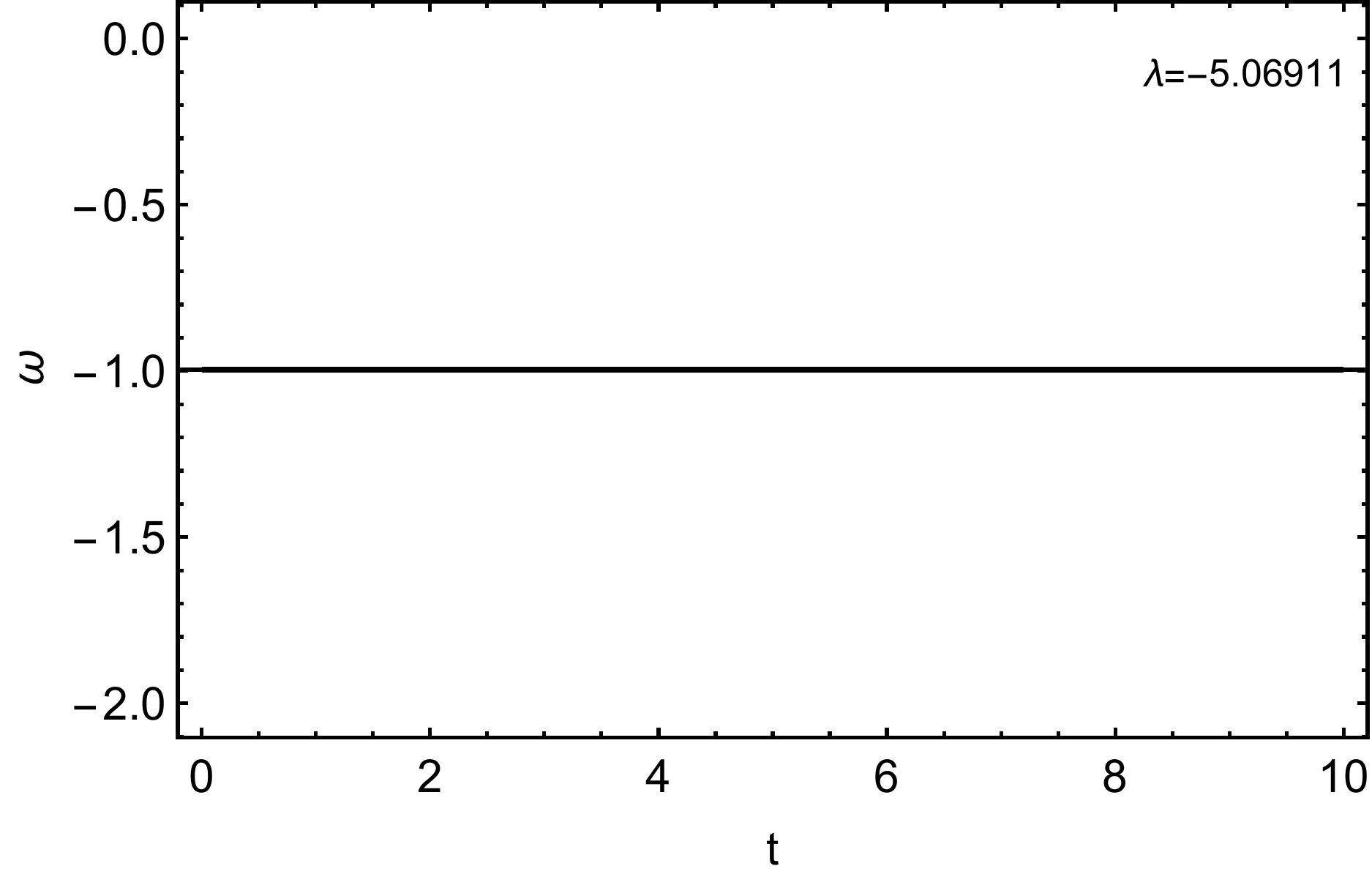}
\caption{Variation of the DE EoS $\omega$ with \textit{t} when $a=b=1, k=15, \lambda=-5.06911$.}
\label{fig:2}       
\end{figure}
Fig. 3(a) shows the variation of the pressure $p$ and energy density $\rho$ when $a=b=1, k=15, \lambda=-5.06911$.
From the graph, it is obvious that the model is experiencing accelerated expansion with negative $p$ and positive $\rho$.  Here, the model evolves with a large $\rho$ and it converges to become constant at late times. This phenomenon is a clear indication of the presence of DE as the present cosmology believes that the late time accelerating universe is due to the dominant and slowly varying or constant DE with negative pressure and positive energy density \cite{5,18,19,20,21,22,23,24}. In order to predict the nature, the graph of $\omega=\frac{p}{\rho}$ which is the DE EoS parameter is plotted in Fig. 4 which shows that $\omega=-1$. Hence, we can sum up that the $f(R,T)$ gravity model we have constructed turns out to be a DE model, DE in the form of  vacuum energy or the cosmological constant. Fig. 3(b) shows the variation of the pressure $p$ and energy density $\rho$ when $a=b=1, k=15, \lambda=-12.5856$. In this case, the model undergoes expansion at an expedited rate with $p$ and $\rho$ both negative. This negative $p$ can be regarded as the indication of the presence of DE. In this scenario too, we can predict that DE in the form of vacuum energy is dominating the model, as predicted by \cite{5}, NED is possible only if the DE is in the form of vacuum energy. Hence, in both the cases, it is fascinating to see that $F(R,T)$ model behaves as a DE (vacuum energy) model. We have not considered the case when $\lambda>0$ as it yields positive pressure which is not reliable in the present scenario. 

\begin{figure}[H]
\centering
\includegraphics[width=0.4\textwidth]{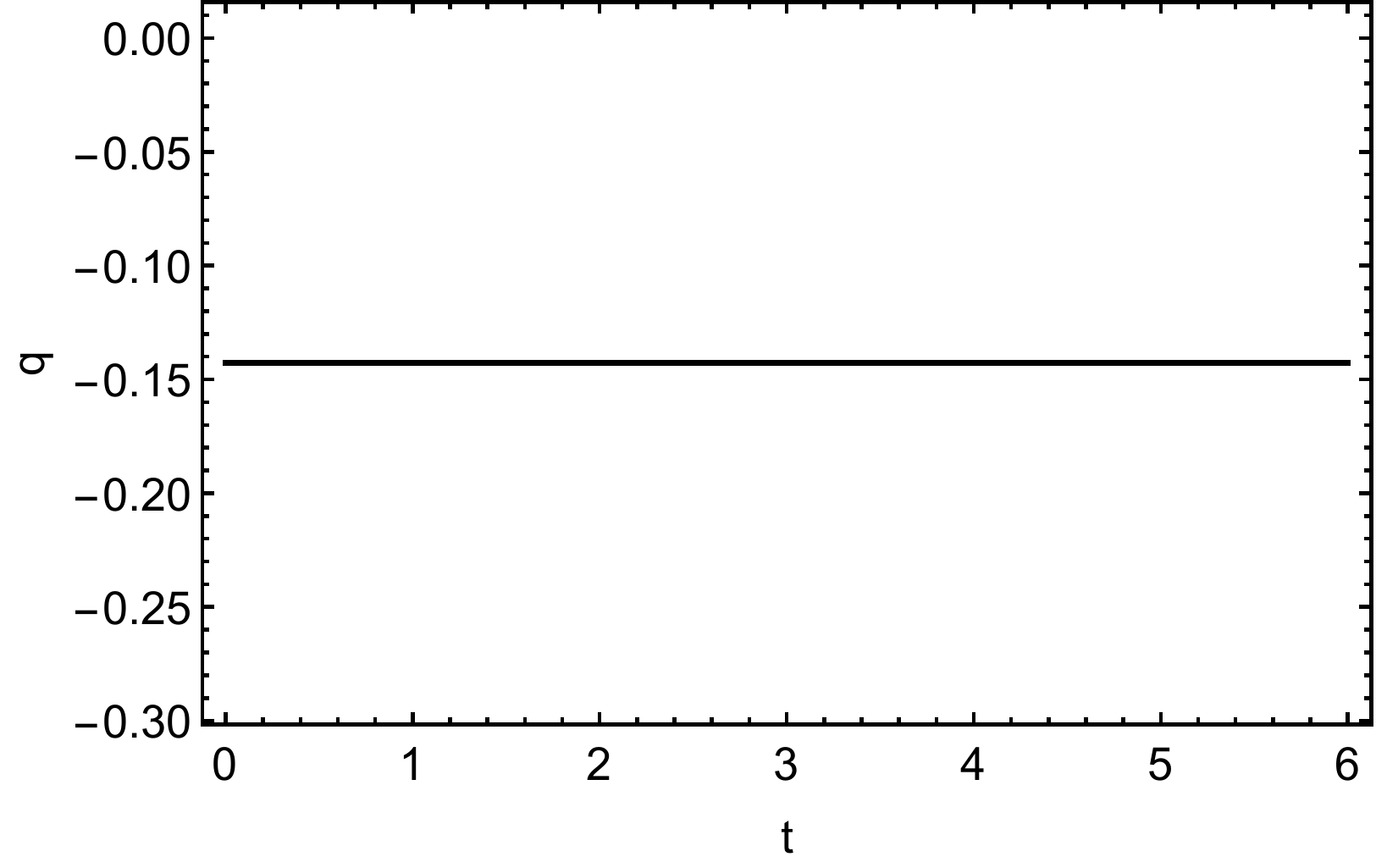}
\caption{Variation of the deceleration parameter $q$ with \textit{t}.}
\label{fig:2}       
\end{figure}
\begin{figure}[H]
\centering
\includegraphics[width=0.4\textwidth]{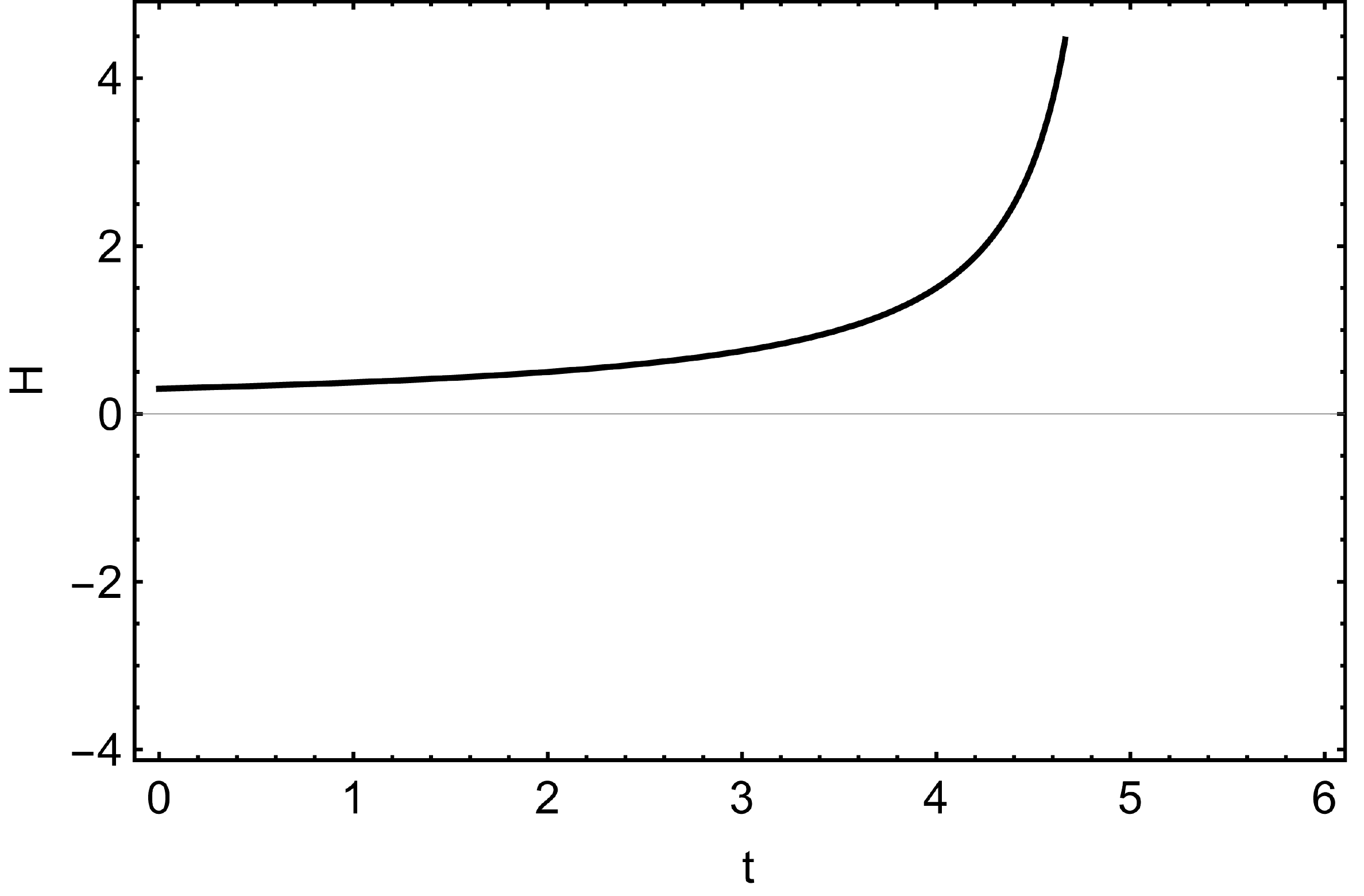}
\caption{Variation of the Hubble's parameter $H$ with \textit{t} when $k=15$.}
\label{fig:2}       
\end{figure}

Accelerated expansion can be attained when $-1<q<0$ whereas $q<-1$ causes super-expansion \cite{72}. Fig. 5 shows that the deceleration parameter $q$ is a negative constant -1.7 all through indicating that the model universe undergoes super-exponential expansion in the entire course of evolution. Fig. 6 shows that the Hubble's parameter $H$ appears to remain almost constant in the early evolution so that our universe was in the inflationary epoch experiencing rapid exponential expansion \cite{73}.

\begin{figure}[H]
\centering
\includegraphics[width=0.4\textwidth]{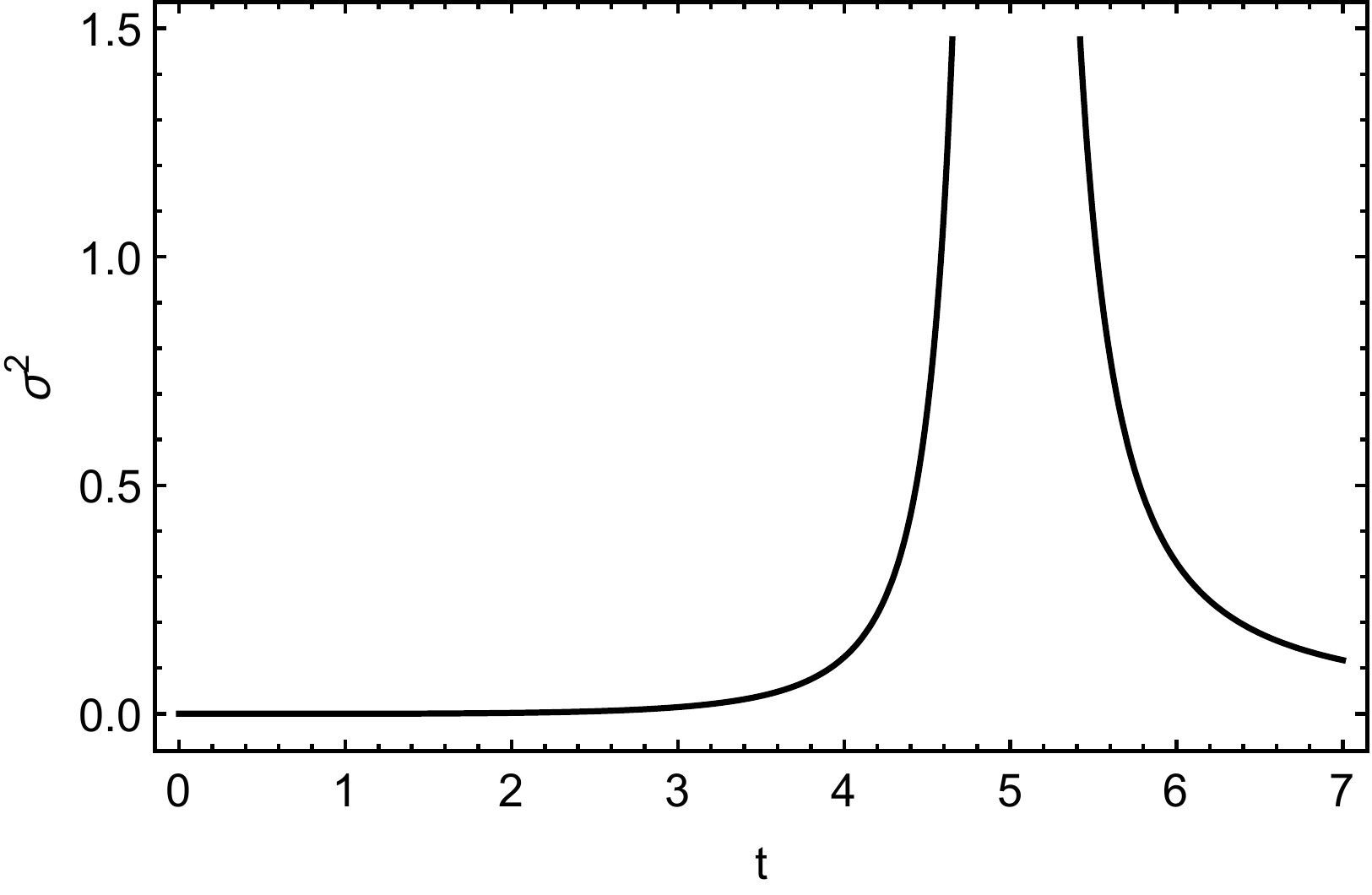}
\caption{Variation of the scalar scalar $\sigma^{2}$ with \textit{t} when $k=15$.}
\label{fig:2}       
\end{figure}

Shear scalar $\sigma^{2}$ provides us the rate of deformation of the matter flow within the massive cosmos \cite{74}. From figure 7, we can see that $\sigma^{2}$ evolves constantly, then diverges after some finite time and again converges to become constant after vanishing for a finite period. From equation (21), the anisotropic parameter $A_h=0$. From these, we can sum up that initially, the isotropic universe expands with a slow and uniform change of shape, but after some finite time, the change becomes faster. Then, the change slows down and tends to become uniform after expanding without any deformation of the matter flow for a finite time period.

\begin{figure}[H]
\centering
\includegraphics[width=0.4\textwidth]{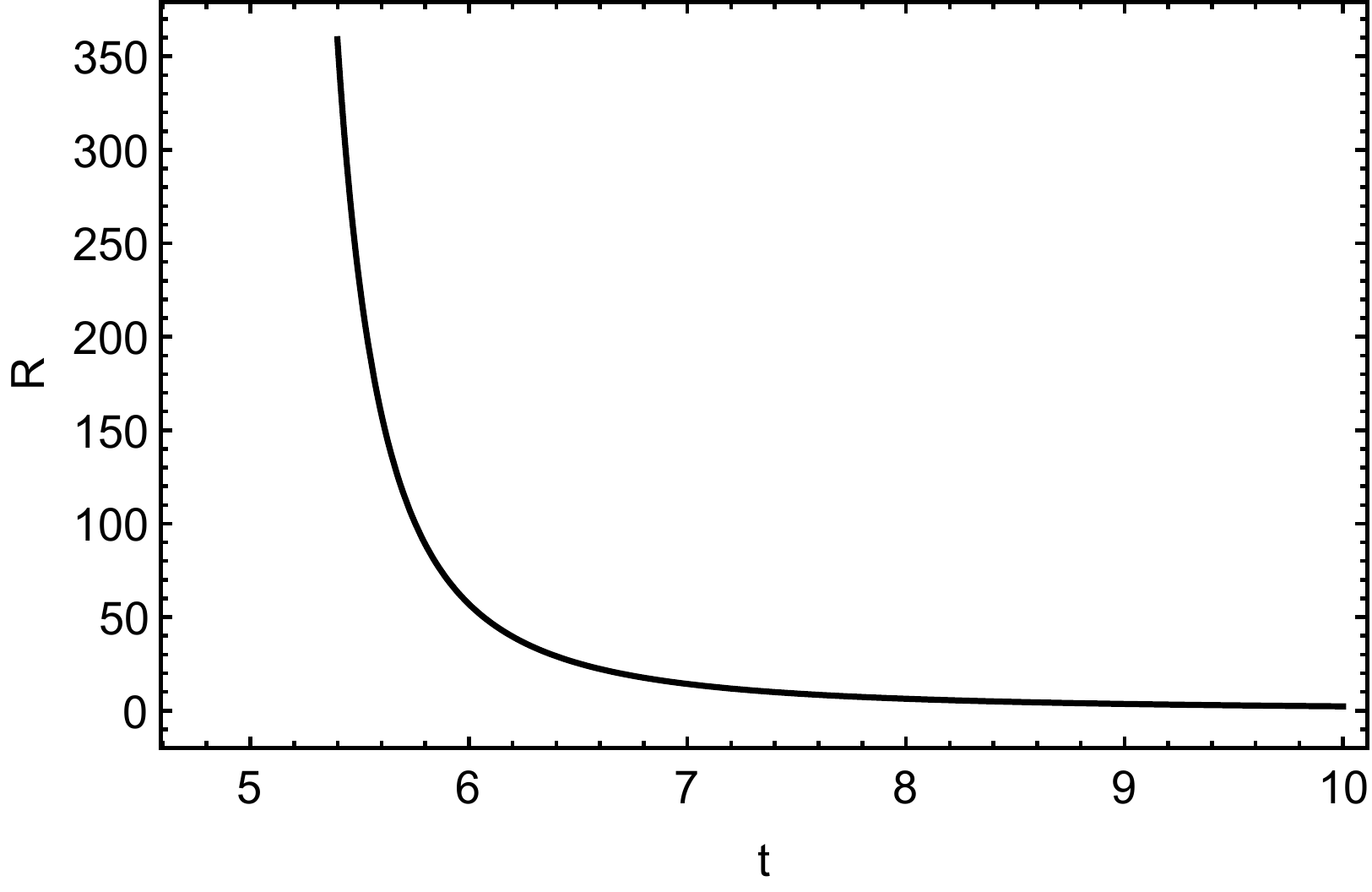}
\caption{Variation of the scalar curvature $R$ with \textit{t} when $k=15$.}
\label{fig:2}       
\end{figure}
Fig. 8 shows the decreasing nature of the scalar curvature $R$ with cosmic time \textit{t}. Similar observation can also be seen in the recent studies \cite{75,76}. At $t=13.8$ Gyr which align with $13.830\pm0.037$ Gyr, the approximate present age of the universe estimated by the latest Planck 2018 result \cite{43}, the scalar curvature is obtained to approach a constant $R=0.72$. $R=0$ corresponds to an exactly flat expanding universe \cite{77,78,79}. However, in the recent years, arguments against the notion of exactly flat universe have been put forwarded by many authors \cite{80,81,82,83}. In the present scenario, the universe is assumed to be close to or nearly flat, but not exactly flat \cite{83,84,85}. Additionally, the latest Planck 2018 results \cite{43} estimating the value of overall density parameter $\Omega$ ranging close to unity can also regarded as an evidence for nearly flat universe, as for an exactly flat universe, $\Omega=1$ \cite{82,85,86}. Hence, our model obtaining a small and constant $R=0.73$ is justified. \cite{87} and \cite{88}, in their studies, assert that $R$ is constant for de-Sitter phase. So, the reason for $R$ approaching a constant can be regarded as an indication for the model approaching the de-Sitter phase dominated by vacuum energy or cosmological constant in the finite time future avoiding singularity. According to \cite{89}, accelerated expansion will lead $R$ to approach a nearly constant value so that the universe behaves in the same manner as a de-Sitter universe in the future. Many other authors have also asserted that the expanding universe will end at the de-Sitter phase dominated by vacuum energy, avoiding singularity. \cite{90,91,92,93,94,95,96,97,98,99,100}.

\section{Conclusions}
Within the framework of $f(R,T)$ gravity, we have analysed a spherically symmetric space-time in 5D setting. We have obtained an isotropic model universe undergoing super-exponential expansion. The variation of the pressure $p$ and energy density $\rho$ with cosmic time $t$ are analysed when $\lambda=-5.06911$ and $-12.5856$. In both the cases, it is fascinating to see that our $F(R,T)$ model behaves as a DE (vacuum energy) model. The model is free from initial singularity. The model expands with a slow and uniform change of shape, but after some finite time, the change becomes faster. Then, the change slows down and tends to become uniform after expanding without any deformation of the matter flow for a finite time period. The scalar curvature $R$ is decreasing with time which is consistent with the recent studies. The model is predicted to approach the de-Sitter phase dominated by vacuum energy or cosmological constant in the finite time future avoiding singularity. We have constructed a model where $f(R,T)$ gravity theory itself behaves as a DE (vacuum energy) model; nonetheless, the work we have put forward is just a toy model. The model needs further deep study considering all the observational findings, which will be our upcoming work.


\begin{thebibliography}{99}

\bibitem{1} A. G. Riess et al., Astron. J. 116 (1998) 1009. 
\bibitem{2} S. Perlmutter et al., Astrophys. J. 517 (1999) 565.
\bibitem{3} M. H. Chan, Phys. Sci. Int. J. 5 (2015) 267.
\bibitem{4} P. J. Peebles,  Ratra B., Rev. Mod. Phys. 75 (2003) 559.
\bibitem{5} S. M. Carroll, Living Rev. Rel. 4 (2001) 1.
\bibitem{6} S. M. Carroll, arXiv:astro-ph/0107571 (2001). 
\bibitem{7} D. Singh, S. Kar, Braz. J. Phys 50 (2020) 673.
\bibitem{8} T. F. Neiser, Adv. Astron 2020 (2020) 8654307.
\bibitem{9} B. Dikshit, Open Astron. 28 (2019) 220.
\bibitem{10} D. Huterer, D. L. Shafer, Rep. Prog. Phys. 81 (2018) 016901.
\bibitem{11} Y. Wang et al., Astrophys. J. Lett. 869 (2018) L8.
\bibitem{12} A. Capolupo, Adv. High Energy Phys. 2018 (2018) 9840351.
\bibitem{13} T. Josset et al., Phys. Rev. Lett. 118 (2017) 021102.
\bibitem{14} S. K. Tripathy et al., Eur. Phys. J. C 75 (2015) 149.
\bibitem{15} M. H. Chan, J. Gravity 2015 (2015) 384673.
\bibitem{16} I. Gontijo, arXiv:1209.1386v2 (2012).
\bibitem{17} S. Alexander et al., Phys. Rev. D 81 (2010) 043511.
\bibitem{18} B. M. Law, Astrophys. Space Sci. 365 (2020) 64.
\bibitem{19} K. P. Singh, P. S. Singh, Chin. J. Phys. 60 (2019) 239. 
\bibitem{20} P. Agrawal et al., Phys. Lett. B 784 (2018) 271. 
\bibitem{21} S. Ray et al., Int. J. Theor. Phys. 52 (2013) 4524. 
\bibitem{22} N. Straumann, Lect. Notes Phys. 721 (2007) 327.
\bibitem{23} J. C. N. d. Araujo, Astropart. Phys. 23 (2005) 279. 
\bibitem{24} P. Wu, H. Yu, Nucl. Phys. B. 727 (2005) 355.
\bibitem{25} A. Ijjas, P. J. Steinhardt, Phys. Lett. B 795 (2019) 666.
\bibitem{26} W. Wong et al., EPJ Web of Conference 206 (2019) 09012.
\bibitem{27} R. J. Nemiroff et al., JCAP 06 (2015) 006.
\bibitem{28} S. Fay, Phys. Rev. D 89 (2014) 063514.
\bibitem{29} I. Sawicki, A. Vikman, Phys. Rev. D 87 (2013) 067301.
\bibitem{30} A. D. L. Macorra, G. German, Int. J. Mod. Phys. D 13 (2004) 1939.
\bibitem{31} W. H. Huang, J. Math. Phys. 6 (1990) 1456.
\bibitem{32} L. Parker, S. A. Fulling, Phys. Rev. D 7 (1973) 2357.
\bibitem{33} S. W. Hawking, G. F. R. Ellis,  The Large Scale Structure of Space-time, Cambridge (England), New York: Cambridge University Press; 1973.
\bibitem{34} C. J. Fewster,  arXiv:1208.5399v1 (2012).
\bibitem{35} N. Graham, K. D. Olum, Phys. Rev. D 67 (2003) 085014.
\bibitem{36} M. Visser, C. Barcelo, arXiv:gr-qc/0001099v1 (2000). 
\bibitem{37} M. J. Pfenning, L. H. Ford L.,  arXiv:gr-qc/9805037v1 (1998).
\bibitem{38} A. D. Helfer, Class. Quant. Grav. 15 (1998) 1169.
\bibitem{39} A. D. Helfer, Mod. Phys. Lett. A 13 (1998) 1637.
\bibitem{40} L. H. Ford, T. A.  Roman, Phys.Rev. D 53 (1996) 5496.
\bibitem{41} T. A. Roman, Phys. Rev. D 33 (1986) 3526.
\bibitem{42} H. Epstein et al., Nuovo Cim. 36 (1965) 1016.
\bibitem{43} P. Collaboration et al.: arXiv:1807.06209v2 (2019).
\bibitem{44} T. Harko et al.,  Phys. Rev. D 84 (2011) 024020.
\bibitem{45} V. R. Chirde, S. H. Shekh, Bulg. J. Phys. 46 (2019) 94.
\bibitem{46} R. Myrzakulov, arXiv:1205.5266v3 (2020).
\bibitem{47} P. Sahoo et al.: doi: 10.1139/cjp-2019-0494 (2020).
\bibitem{48} M. S. Singh, S. S.  Singh, New Astronomy 72 (2019) 36.
\bibitem{49} D. D. Pawar et al. J. Astrophys. Astr. 40 (2019) 0013.
\bibitem{50} R. Zia et al., Int. J. Geom. Methods Mod. Phys. 15 (2018) 1850168.
\bibitem{51} M. Srivastava, C. P. Singh, Astrophys. Space Sci. 363 (2018) 117.
\bibitem{52} V. Fayaz et al.,  Eur. Phys. J. Plus 131 (2016) 22.
\bibitem{53} G. Sun, Y. C. Huang, Int. J. Mod. Phys. D 25 (2016) 1650038.
\bibitem{54} R. K. Mishra et al., Int. J. Theor. Phys. 55 (2016) 1241.
\bibitem{55} C. P. Singh, P. Kumar, Astrophys. Space Sci. 361 (2016) 157.
\bibitem{56} M. J. S. Houndjo, O. F. Piattella, Int. J. Mod. Phys. D 21 (2012) 1250024.
\bibitem{57} B. Mishra et al., Adv. High Energy Phys. 2016 (2016) 8543560.
\bibitem{58} M. Zubair et al., Can. J. Phys. 94 (2016) 1289.
\bibitem{59} N. Ahmed et al.: NRIAG J. Astron. Geophys. 5 (2016) 35.
\bibitem{60} V. U. M Rao, D. C. P. Rao, Astrophys. Space Sci. 357 (2015) 65.
\bibitem{61} M. Jamil et al., Eur. Phys. J. C 72 (2012) 1999.
\bibitem{62} M. J. S. Houndjo, Int. J. Mod. Phys. D 21 (2012) 1250003.
\bibitem{63} T. Kaluza, Sitzungsber. Preuss Akad. Wiss. Berlin Math. Phys. K1 (1921) 966.
\bibitem{64} O. Klein, Z. Phys. 37 (1926) 895.
\bibitem{65} S. K. Banik, K. Bhuyan,  Pramana J. Phys. 88 (2017) 26.
\bibitem{66} G. P. Singh et al., Pramana J. Phys. 63 (2004) 937.
\bibitem{67} W. J. Marciano, Phys. Rev. Lett. 52 (1984) 489. 
\bibitem{68} Z. E. Alvax, M. B. Gavela, Phys. Rev. Lett. 51 (1983) 931.
\bibitem{69} A. H. Guth : Phys. Rev. D 23 (1981) 347.
\bibitem{70} S. Chakraborty, U. Debnath, Int. J. Theor. Phys. 49 (2010) 1693.
\bibitem{71} T. Koivisto, Class. Quant. Grav. 23 (2006) 4289.
\bibitem{72} G. P. Singh, B. K. Bishi : Adv. High Energy Phys. 2017 (2017) 1390572.
\bibitem{73} G. U. Crevecoeur, arXiv:1603.06834v2 (2017).
\bibitem{74} G. F. R. Ellis, H. V  Elst. NATO Adv. Study Inst. Ser. C. Math. Phys. Sci. 41 (1999) 1 .
\bibitem{75} P. Pavlovic, M. Sossich, Phys. Rev. D 95 (2017) 103519.
\bibitem{76} E. A. Pa shitskii, V. I. Pentegov, J. Exp. Theor. Phys. 122 (2016) 52.
\bibitem{77} V. G. Gueorguiev, A. Maeder, Universe 6 (2020) 108.
\bibitem{78} M. Kleban, L. Senatore, JCAP 10 (2016) 022.
\bibitem{79} J. J. Bevelacqua, Fizika A (Zagreb) 15 (2006) 133.
\bibitem{80} E. D. Valentino et al., Nat. Astron. 4 (2020) 196.
\bibitem{81} W. Javed et al., Eur. Phys. J. C 80 (2020) 90.
\bibitem{82} M. Khodadi et al., Eur. Phys. J. C 75 (2015) 590.
\bibitem{83} G. G. L. Nashed, W. E. Hanafy, Eur. Phys. J. C 74 (2014) 3099.
\bibitem{84} R. J. Adler, J. M. Overduin, Gen. Relativ. Gravit. 37 (2005) 1491.
\bibitem{85} J. J. Levin, K. Freese, Nucl. Phys. B 421 (1994) 635.
\bibitem{86} M. Holman, Found. Phys. 48 (2018) 1617. 
\bibitem{87} R. K. Tiwari, J. Phys. Conf. Ser. 718 (2016) 032009.
\bibitem{88} Y. Kim et al., J. Korean Phys. Soc. 42 (2003) 573. 
\bibitem{89} K. Falls et al.: Class. Quantum Grav. 35 (2018) 135006.
\bibitem{90} S. Basilakos et al., Eur. Phys. J. C 78 (2018) 684.
\bibitem{91} S. Carneiro, Int. J. Mod. Phys. D 15 (2006) 2241. 
\bibitem{92} I. Dymnikova, Universe 5 (2019) 111.
\bibitem{93} L. Dyson et al., JHEP 10 (2002) 011.
\bibitem{94} S. Nojiri S, S. D. Odintsov, Phys. Lett. B, 595 (2004) 1.
\bibitem{95} L. M. Krauss, G. D. Starkman, Astrophys. J. 531 (2000) 22.
\bibitem{96} T. Markkanen, Eur. Phys. J. C 78 (2018) 97.
\bibitem{97} A .D. Sakharov, Sov. Phys. JETP 22 (1966) 241.
\bibitem{98} A. A. Starobinsky, Grav. Cosmol. 6 (2000) 157. 
\bibitem{99} G. J. M. Zilioti  et al., Adv. High Energy Phys. 2018 (2018) 6980486.
\bibitem{100} Y. Zhang et al., Adv. High Energy Phys. 2020 (2020) 7263059. 

\end{thebibliography}
\end{document}